\documentclass[10pt]{article}
\usepackage{dcolumn}
\usepackage{bm}
\usepackage{verbatim}       

\usepackage{graphicx}

\usepackage{amsthm}
\usepackage{amssymb}

\usepackage{bm}
\usepackage{amstext}
\usepackage{psfrag}
\usepackage{amsmath}
\usepackage{amsfonts}
\usepackage{units}
\usepackage{mathrsfs}

\newcommand{\dd}{{\rm d}}

\newcommand{\bd}{\begin{definition}}                
\newcommand{\ed}{\end{definition}}                  
\newcommand{\bc}{\begin{corollary}}                 
\newcommand{\ec}{\end{corollary}}                   
\newcommand{\bl}{\begin{lemma}}                     
\newcommand{\el}{\end{lemma}}                       
\newcommand{\bp}{\begin{proposition}}            
\newcommand{\ep}{\end{proposition}}                
\newcommand{\bere}{\begin{remark}}                  
\newcommand{\ere}{\end{remark}}                     

\newcommand{\bt}{\begin{theorem}}
\newcommand{\et}{\end{theorem}}

\newcommand{\be}{\begin{equation}}
\newcommand{\ee}{\end{equation}}

\newcommand{\bit}{\begin{itemize}}
\newcommand{\eit}{\end{itemize}}
\newtheorem{theorem}{Theorem}[section]
\newtheorem{corollary}[theorem]{Corollary}
\newtheorem{lemma}[theorem]{Lemma}
\newtheorem{proposition}[theorem]{Proposition}
\theoremstyle{definition}
\newtheorem{definition}[theorem]{Definition}
\theoremstyle{remark}
\newtheorem{remark}[theorem]{Remark}

\usepackage{wrapfig}

\begin{document}

\title{Hints at a relationship between friction and relativistic physics\footnote{Talk  at the conference DICE2022, Spacetime - Matter - Quantum Mechanics `Quantum riddles and spacetime oddities', Castiglioncello, Sep.\ 19-23,  2022.}}

\author{E. Minguzzi\thanks{
Dipartimento di Matematica Applicata ``G. Sansone'', Universit\`a
degli Studi di Firenze, Via S. Marta 3,  I-50139 Firenze, Italy.
E-mail: ettore.minguzzi@unifi.it} }

\date{}

\maketitle

\begin{abstract}
\noindent In applied mechanics Reye's law (1860) establishes, via energy arguments, that the mass  of the debris produced by dry friction  in the contact of rigid bodies is proportional to the work done by  friction forces. This result has long been used for the  determination of the distribution of pressure  in the contact of rigid bodies, and hence for the design of brakes. In this work I show that, when bodies losing mass due to friction are treated, as they should, as variable mass systems, a relationship analogous to the relativistic mass formula is recovered. This result suggests that mathematical structures typical of relativistic physics could have been discovered prior to 1905, without making any reference to electromagnetism, group theory or the speed of light. Also this result could point to the existence of a physical theory depending on two constants, the speed of light (Reye's constant) and a   universal frictional deceleration with respect to an absolute frame. The limit of the theory for vanishing friction would give Special Relativity, as the absolute frame would become unobservable, while the limit for the speed of light going to infinity would lead to Aristotelian mechanics, i.e.\ a classical mechanics type theory presenting universal friction. Finally, I present a reference frame transformation that displays these features and I apply the theory to some open cosmological problems.
\end{abstract}

\section{Introduction}

In the last two decades I have been working on the foundations of relativity theory and causality \cite{minguzzi18b}, while teaching courses in Rational (Analytical) Mechanics to students in mechanical engineering.
I enjoyed teaching these courses as I learned a lot particularly exploring results in mechanics that do not typically belong to the curriculum of a physicist. This work is the result of one such exploration in which my two interests in the foundations of relativity physics and in mechanics went to interact.

As a physicist I have always been puzzled by the following question: given two rigid bodies sharing a flat surface, how is the pressure between them going to distribute? If the normal force is $N$ and the area of contact is $A$, it could be tempting to answer: $p=N/A$. However, this looks just as the simplest possible answer though not necessarily the correct one. For instance, should we expect this answer to change if the two bodies are in relative motion? Should the pressure depend on the relative velocity at the  contact point? At first, I really did not think much about this question because I supposed that a correct answer would necessarily rely on a study of the deformation of the  bodies and hence on elements of elasticity theory. Only in this way, I thought, one could really elaborate some argument leading to the justification of some  pressure distribution.

My arguments appeared to be correct but were in fact, as I would later realize,  quite wrong. The problem of finding the distribution of pressure had been already solved in 1860 by the German mathematician Reye through an ingenious  argument that involved the consideration of friction and that completely disregarded elasticity \cite{reye60}. His argument is elegant and marries well with the idealized notion of rigid body that is so much used in mechanics. There is really no need to abandon the rigid body idealization just to solve the pressure distribution problem. \\
%

\begin{wrapfigure}{r}{0.5\textwidth}
  \begin{center}
    \includegraphics[width=0.30\textwidth]{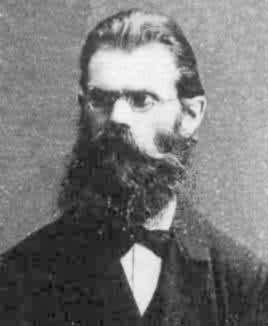}
  \end{center}
  \caption{Karl Theodore Reye}
\end{wrapfigure}
A historical note: 
Karl Theodore Reye (1838–1919) was a German
mathematician who contributed significatively to
projective geometry. \\
He studied at the polytechnic in
Hanover and in Z\"urich, where he turned to theoretical
physics under the influence of Rudolf Clausius.
When he was 22, in 1860, he published
the landmark paper \cite{reye60} “On the theory of pivot
friction” where he established a successful theory
of wear under friction.\\

In his study Reye considers the friction of two bodies $A$ and $B$, sliding one against the other
on a common surface $\Sigma$ under friction. One of the bodies, say $A$, is considered
hard so that friction wears out just the (soft) body $B$.
It can be assumed that a given proportion of the work done by friction forces goes in the breaking of the intermolecular bonds and hence in the detachment of mass from the soft body.
He then  establishes \cite{reye60} \\

{\bf Reye’s wear law (1860)}:
The wear, that is the mass of the debris produced, is proportional
to the work done by friction forces:
\begin{equation} \label{rey}
m_{debris} =\frac{1}{c^2}
\vert L_f \vert
\end{equation}
where $c$ is Reye’s constant (it has dimensions of a velocity and it depends on the type
of materials at contact). He also showed how to use this formula to deduce the pressure
distribution on the contact surface via geometrical arguments (we shall see some examples in a moment, uniform pressure turns out to be untenable). The experimental verification of the law was sufficiently good, in fact resistance to wear  was defined and measured based on its validity \cite{giovannini35}.

Opatowski \cite{opatowski42} wrote in 1942:

\begin{quote}
The importance of wear by friction is, in some mechanical organs, particularly
brakes, so great, that the ideas which Reye published 80 years ago
still form today the theoretical basis for their design.
\end{quote}


Although there were much posterior derivations of Eq.\ (\ref{rey}), most notably by Archard \cite{archard53} (1953), or by  M. M. Khrushchov and  M. A. Babichev (1960) for  the metal-metal contact, these were actually derivations based on other laws valid in specific contact models. It is no surprise that the same law (\ref{rey}) can be reached starting from several different microscopical mechanisms, based on microscopic laws that apply only in specific circumstances.
The strongest argument seems Reye's as it is based on energy considerations alone and as is, for this reason, universal, i.e.\ independent of specific microscopic models.


In Italy Reye's law, also called Reye's assumption, entered very early in  the program of the university course ``Meccanica applicata alle macchine" (Mechanics of machines) and has always been a traditional topic of study, particularly for its usefulness in the design of brakes.

Villaggio \cite{villaggio01} states that Reye's work was first introduced in university courses in Italy by  Modesto Panetti. Panetti wrote a paper on applications of Reye's law in 1908 \cite{panetti08} while first editions of his university book \cite{panetti47} introducing Reye's law to students in engineering were published in 1913. However, I found that this account is not completely accurate.
The teaching of   ``Meccanica applicata alle macchine" started earlier in the ``Scuola d'applicazione per gli ingegneri" established in Turin in 1860. The course was first thought by Prospero Richelmy,  from the foundation of the school to 1880, then by Ferdinando Zucchetti till 1882 and then by Scipione Cappa assisted by Elia Ovazza  for several years \cite{curioni84}. It was Cappa who introduced in his 1890 university book  Reye's law \cite{cappa90} devoting to it considerable space (also cited in \cite{ovazza90}). As the Italian tradition of ``Meccanica applicata alle macchine'' originated from  this school, it appears likely that most Italian University professors, including Panetti, learned of Reye's work from Cappa's 1890 book and posterior editions. Some early references dealing with Reye's law are \cite{pomini22,rubino44}, while as  more recent references \cite{ferrari66,funaioli73} can be mentioned.

 Early sources in English  that mention Reye's work are certainly less numerous and include the 1891 translation `The Constructor' by Henry Harrison Suplee of F. Reuleaux classic, or Weisbach book \cite{weisbach70}.

Reye's law admits an infinitesimal formulation. Denoting with $m$ the mass of the soft body $B$, with $W_f$ the power of the friction
force (necessarily negative), with ${\bm F}_f$ the friction force and with ${\bm v}$ velocity of $B$ relative to $A$,
\[
\dot m=\frac{1}{c^2} W_f=\frac{1}{c^2} {\bm F}_f\cdot {\bm v}= - \frac{1}{c^2} F_f v ,
\]
where we used the fact that the friction force is oriented opposite to the relative velocity.

The  power of friction forces might depend on the region of the contact surface $\Sigma$ considered. Thus the body B might wear out more at some points than others.
The previous formula has to be applied on a small region of $\Sigma$ of area $\Delta A$, then $F_f= \mu p \Delta A$, where $\mu$ is the friction coefficient and $p$ is the pressure. Denoting with $\rho$ the density, we have $\dd m= - \rho \Delta A \dd h$, where $h$ is the width of the wear out part. As a result
\[
\dot h= \frac{\mu}{ \rho } \frac{1}{c^2} p v.
\]
This equation shows that wear is proportional to the product of pressure and velocity at the contact point. Geometrical considerations impose specific forms  for the function $h:\Sigma \to \mathbb{R}$ and hence for the product $pv$. This ultimately allows the determination of the pressure distribution.

\begin{figure}[h!]
\centering
 \includegraphics[width=10cm]{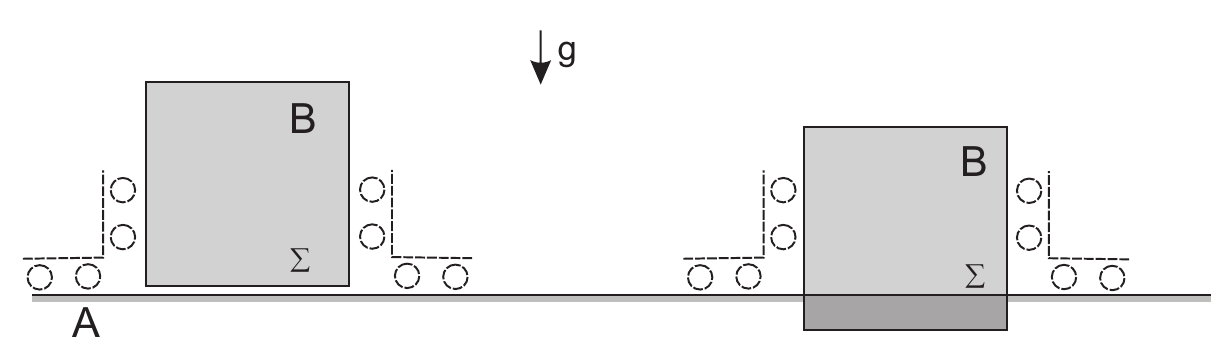}
 \caption{A block loses mass due to friction while it moves. The block is aided by an idealized device that avoids its tilting. As a result pressure is uniformly distributed as $\dot h$ does no depend on the contact point. } \label{stra}
\end{figure}

\begin{figure}[h!]
\centering
 \includegraphics[width=10cm]{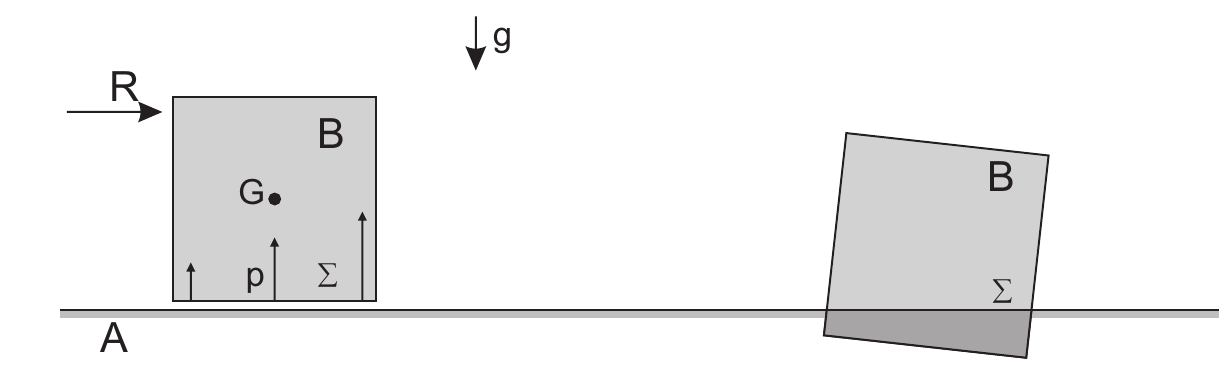}
 \caption{A block loses mass due to friction while it moves. As it is pushed by a force $R$, and the angular velocity is negligible, the pressure distributes so as to balance the mechanical momentum with respect to $G$. This results in an uneven pressure distribution and hence in the (slow) tilting of the block. In first approximation the pressure distribution must be linear,  $p=a+b x$, as  $h$ varies linearly with the horizontal coordinate $x$.} \label{stra2}
\end{figure}

For instance, Figures \ref{stra} and   \ref{stra2} show what happens to a block that loses mass due to wear depending on different geometrical constraints imposed on it.

\begin{figure}[ht!]
\centering
 \includegraphics[width=9cm]{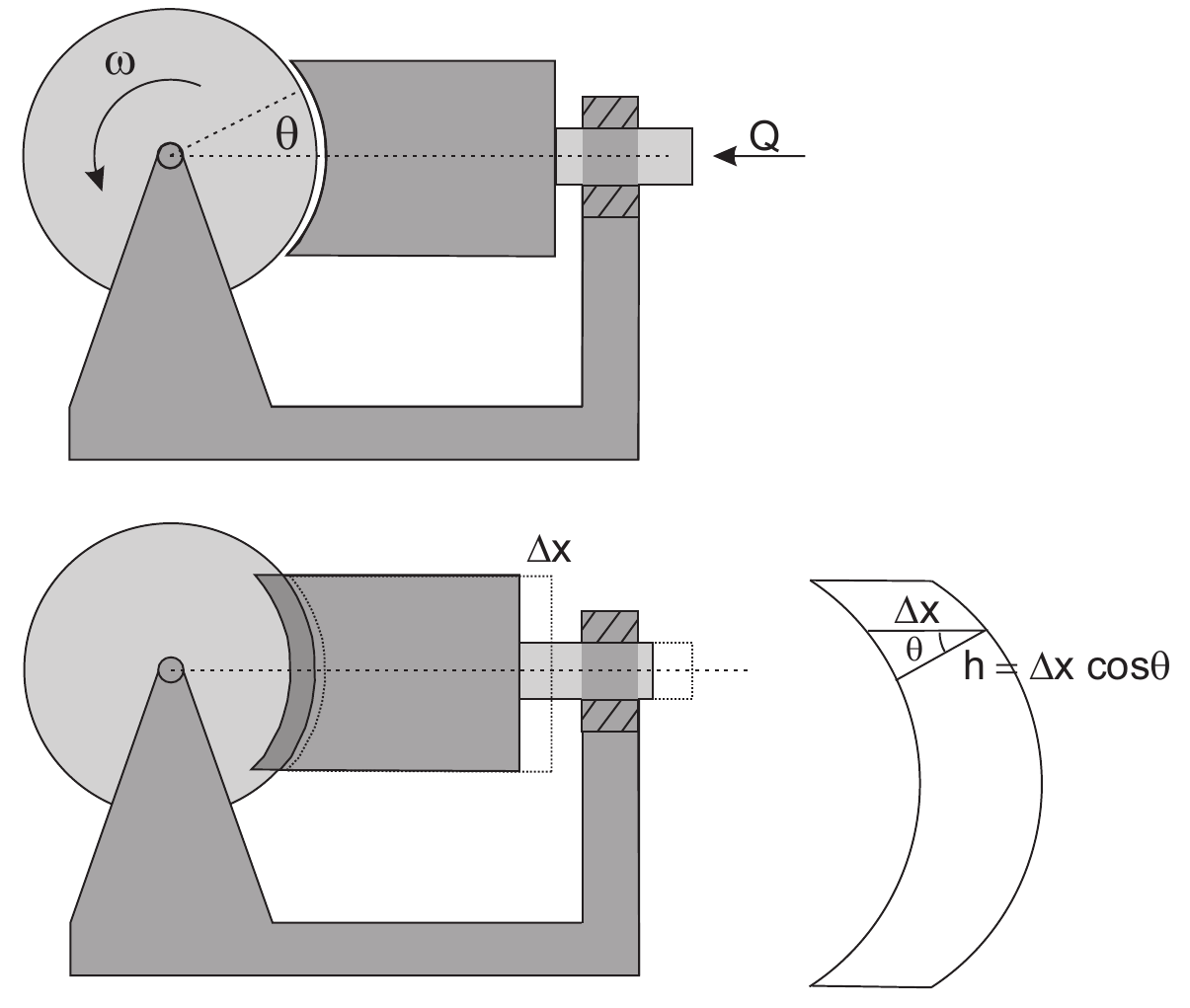}
 \caption{A simple brake in which, by Reye's law, $p=a \cos \theta$, for some constant $a>0$.} \label{cep}
\end{figure}

An interesting design of a brake is depicted in Figure \ref{cep}. Here the hard part $A$ is the wheel, while the soft part $B$ is the  block that is  pushed horizontally towards the wheel by a force ${\bm Q}$. As the wheel spins with angular velocity $\omega$ the relative velocity at each contact point might depend on time but  is actually independent of the point and hence of $\theta$. As a result the pressure is proportional to $\dot h$. The geometrical constraint, accurately show in the figure, shows how the block wears as it translates. The result is
 $\dot h \propto \cos \theta$ which implies $p=a\cos \theta$ for some constant $a$. The constant $a$  could be determined by using the range of $\theta$ and the friction coefficient $\mu$ by imposing that the contact force between wheel and block  equals $Q$.

\begin{figure}[ht!]
\centering
 \includegraphics[width=10cm]{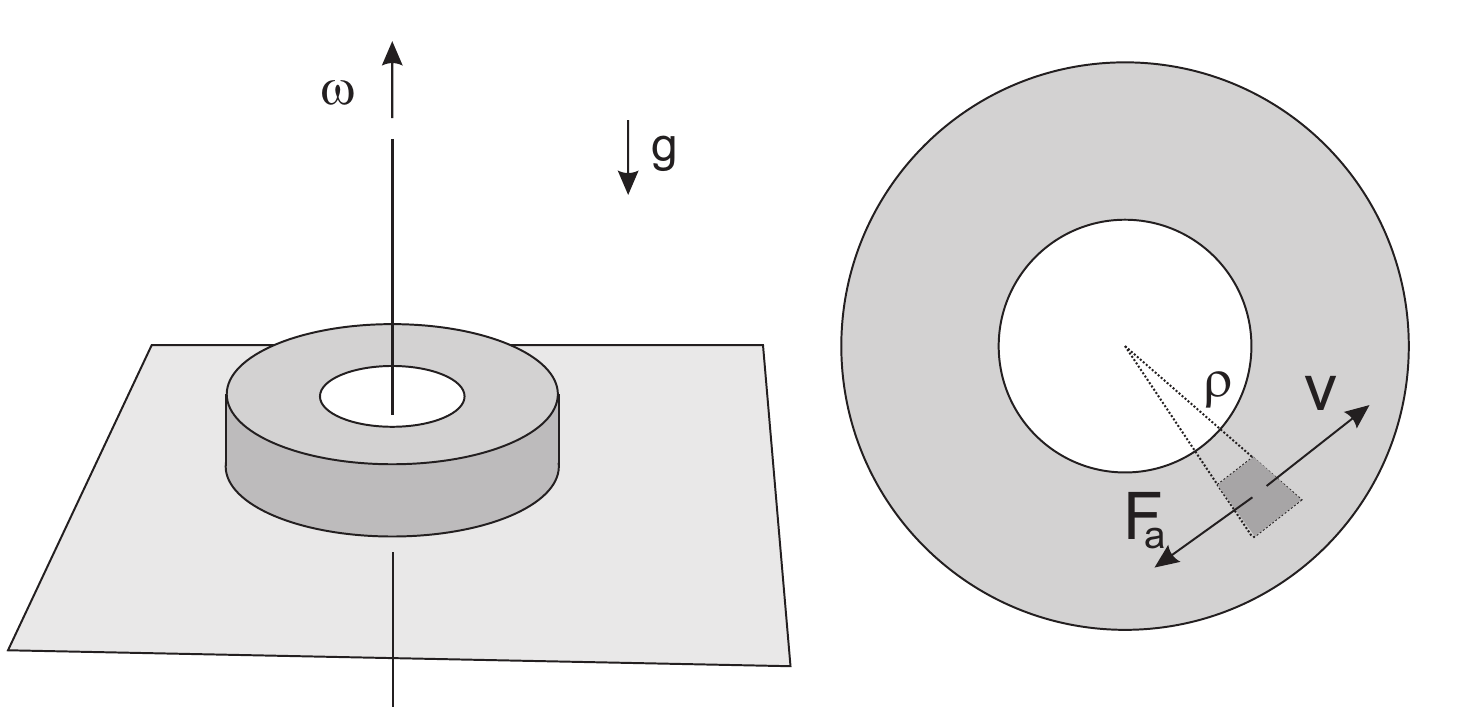}
 \caption{The surface of contact $\Sigma$. Friction and relative velocity depend on the region considered.} \label{disa}
\end{figure}

\begin{figure}[ht!]
\centering
 \includegraphics[width=10cm]{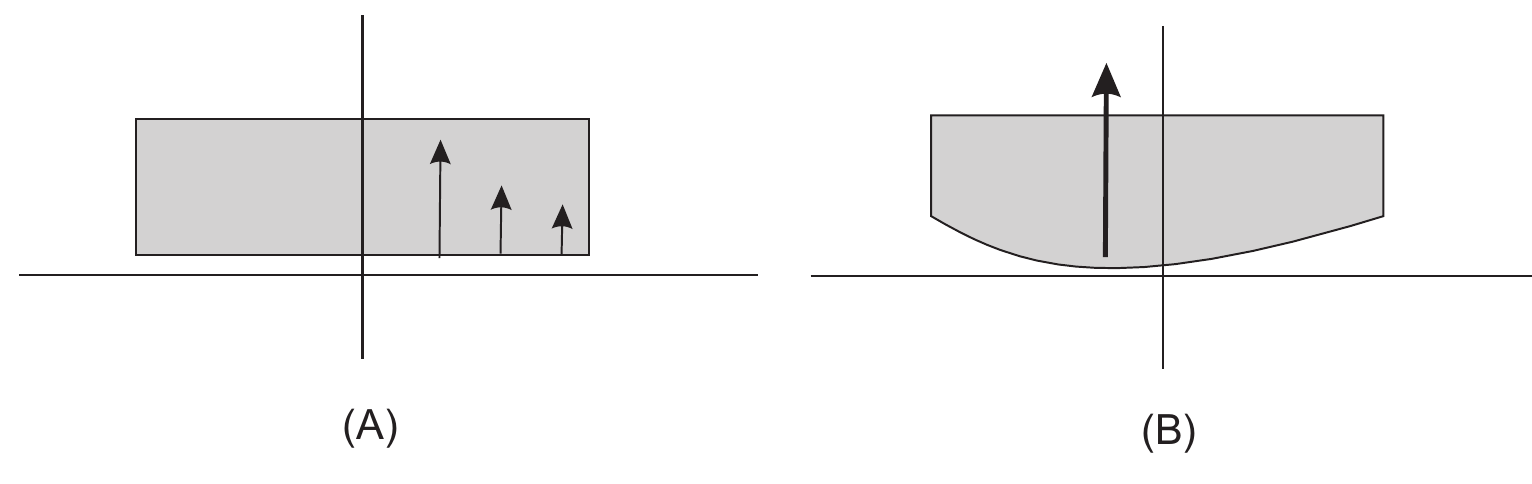}
 \caption{Inhomogeneous wearing (B) naturally leads to more load and friction in those regions that have previously lost less mass, and that are, as a result, more exposed. Ultimately, the stationary configuration is that of homogeneous wearing as illustrated in figure \ref{dis}.} \label{disb}
\end{figure}

\begin{figure}[ht!]
\centering
 \includegraphics[width=10cm]{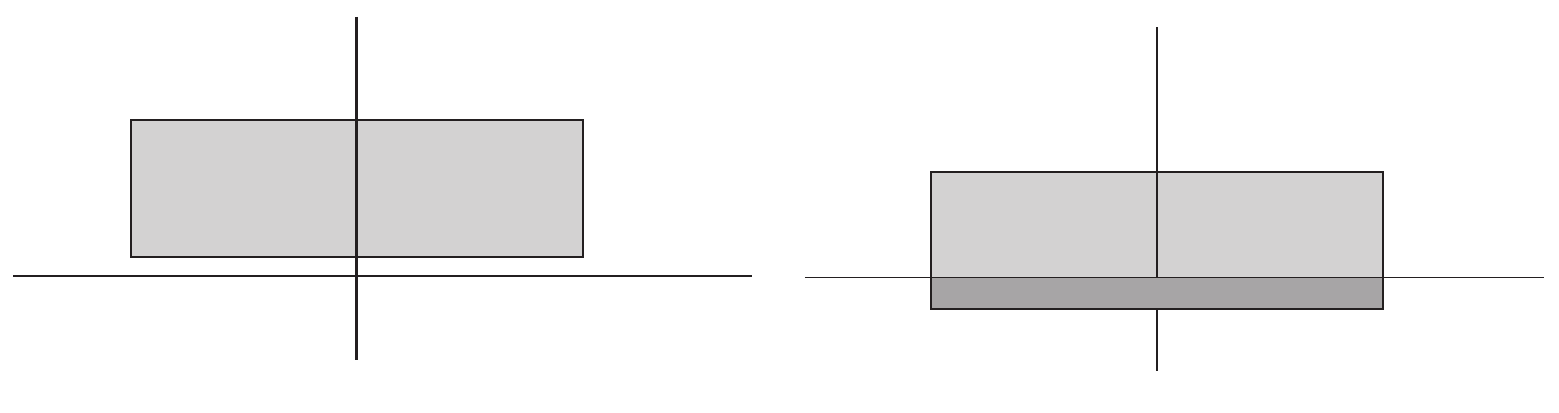}
 \caption{Homogeneous wearing follows from the request that the axis does not tilt.} \label{dis}
\end{figure}

As a further example we can consider a  cylindrical annulus  of mass $m$ spinning over a flat horizontal surface (Fig.\ \ref{disa}). The interior radius of the annulus is $r$ while the exterior radius is $R$. If the axis of the cylinder does not tilt as it wears (we might assume that there is some mechanism preventing it from happening), the function $\dot h$ is constant (Fig.\ \ref{dis}). Indeed, if the cylinder were to wear in an inhomogeneous way that would lead to more friction precisely in those regions that are less worn and that are therefore more exposed to contact and friction (Fig.\ \ref{disb}). We conclude that the product of pressure and velocity is constant, and since the latter is proportional to the distance $\rho$ from the axis,  $p=a/\rho$ for some constant $a$. From here it is easy to calculate  that $a=\frac{ m g}{2\pi(R-r)}$.

\section{Variable mass system}

So far we were concerned with the problem of the pressure distribution and we assumed, as it is reasonable in practical problems, that the loss of mass is negligible when it comes to consider the dynamics of a body.
Let us study the dynamical problem by assuming that Reye's law is exact, also well beyond the regimes of velocity in which it has been verified, and let us solve the dynamical equations with all rigor, by using the laws of classical mechanics.

\begin{figure}[ht!]
\centering
 \includegraphics[width=11cm]{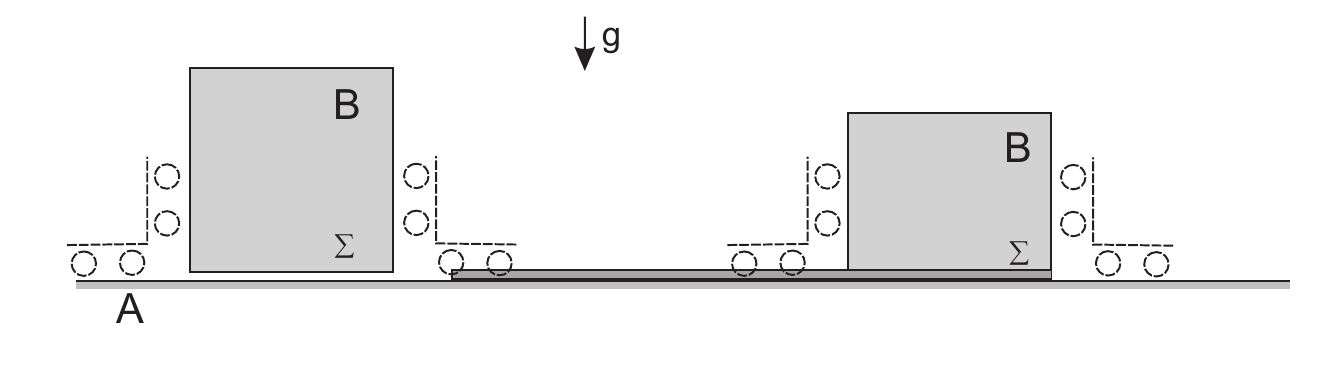}
 \caption{The block loses mass. The debris come to rest in the reference frame of the horizontal plane (hard body $A$).} \label{dism}
\end{figure}

Rigorously speaking, a body that loses mass should be treated as a variable mass system and hence the first cardinal equation of dynamics should be replaced by the  Painlev\'e and Seeliger formula
(1890). This formula  follows from the conservation of momentum and is familiar
from rocket theory \cite{irschik04}. In the block example we shall write (see Fig.\ \ref{dism})
\begin{equation} \label{nhi}
m \dot{\bf v}= {\bf F}_f-\dot{m} ({\bf v}-{\bf v}_d)
\end{equation}
where ${\bf v}_d$ is the velocity of the debris after detachment. In our case the debris do not move with respect to the hard body $A$ (the horizontal plane) and so  ${\bf v}_d=0$.

Denoting with ${\bf p}:=m {\bf v}$ the linear momentum we obtain the
system of equations
\begin{align}
\dot{m}&= \frac{1}{c^2}{\bf F}_f \cdot {\bm v} =-\frac{1}{c^2}\, a_f p,  \quad \textrm{(Reye)} \label{jhg}\\
\dot{{\bf p}}&= {\bf F}_f=-a_f m \hat{{\bf v}}. \quad \textrm{(Painlev\'e and Seeliger)} \label{jhh}
\end{align}
where we used the formula for Coulomb's dynamical friction ${\bf F}_f =-\mu m g \hat{{\bf v}} =-m a_f \hat{{\bf v}}$ and where we introduced the constant $a_f:=\mu g$, which is a characteristic acceleration related to friction. Actually many of the below calculations generalize to  $a_f$ dependent on velocity, and hence to Stokes friction.

\subsection{A puzzling result}

Multiplying Eq.\ (\ref{jhg}) by $m$, multiplying scalarly Eq.\ (\ref{jhh}) by $\frac{1}{c^2}\, {\bf p}$, and taking the
difference we obtain that there is a constant $\tilde m$ such that
\begin{equation} \label{kos}
m^2 -\frac{1}{c^2} {\bf p}^2 =\tilde m^2 ,
\end{equation}
or equivalently
\begin{align}
m&=\frac{\tilde m}{\sqrt{1-\frac{v^2}{c^2}}}, &v<c   \qquad \ \quad \qquad if \ \tilde m>0, \label{ngx}\\
mc^2&=p \,c,  &v=c, \qquad \ \quad \qquad if \
\tilde m=0, \label{ngy}\\
m&=\frac{\vert \tilde m\vert }{\sqrt{\frac{v^2}{c^2}-1}}, &v>c \qquad if
\ \tilde m \textrm{ is imaginary}, \label{ngz}
\end{align}
Observe that both $v(t)$ and $m(t)$ are time dependent, due to the presence of friction and wear. The just obtained result tells us that they maintain a relationship in time.

As we can see, such relationship is  formally coincident with the relativistic mass formula, where in the parallelism Reye's constant $c$ corresponds to the speed of light. The three cases might be called bradyonic, luxonic and tachyonic.

The physical interpretation of $\tilde m$ in the bradyonic case can be read off the above formula, that is,
$\tilde m$ is the mass for $v \to 0$, namely the mass that the
block will have after wearing by friction is completed, i.e.\ after the block has come to  rest with respect to the frame of the horizontal plane.

From the previous equations, taking into account that the
mass $m$ is always positive, we obtain
\begin{equation} \label{ndf}
\dot{\bf v}= ({\bf F}_f-\dot{m} {\bf v})/m=-{a_f}
(1-\frac{v^2}{c^2})\hat{{\bf v}}.
\end{equation}
This equation  shows that if the body has velocity $c$ it   keeps
moving in straight line preserving it (luxonic case). At that speed the block velocity would be basically insensitive to friction. However, mass and linear momentum  decrease (recall that $p=m v$)
\begin{equation} \label{bsc}
p(t)=p(0)\, e^{-\frac{1}{c}  a_f t}, \qquad \textrm{ for }
v=c.
\end{equation}

In the bradyonic casee Eq.\ (\ref{ndf}) can be integrated and gives for $a_f\ne 0$
\begin{align} \label{kde}
v(t)&=c \tanh (\varphi-\frac{a_f}{c} \,t), & \textrm{for} \ 0\le v<c,
\end{align}
where $\varphi$ is an integration constant (interpreted as the
rapidity) related to the initial velocity ($c \tanh \varphi$). We observe that for the chosen Coulomb friction the time needed to reach a complete stop
is finite, namely $t_f=c \varphi/a_f$.

The dependence of mass on time is
\begin{align}
m(t)&=\tilde m \cosh (\varphi-\frac{a_f}{c}\, t), & \textrm{for bradyonic  case}.
\end{align}
It becomes equal to the  mass $\tilde m$
when the body comes to rest in $A$.

Let us integrate once more in the bradyonic case
\begin{equation} \label{kdr}
x(t)=-\frac{c^2}{a_f} \ln \cosh (\varphi-\frac{a_f}{c} \,t)+x_f=-\frac{c^2}{a_f} \ln \cosh \Big(\frac{a_f}{c}(t_f- t)\Big)+x_f
\end{equation}
where $x_f$ is the coordinate of the stopping point and $t_f$ is the stopping time.


The above `relativistic mass formula' (\ref{ngx}) follows just from the principles of ordinary classical mechanics and is
certainly correct for small $v$, as Reye’s law holds in that limit. Light,
electromagnetism and related concepts did not enter the previous derivation. We might say that pre-1905 ordinary classical mechanics might itself suggest to investigate theories which embody mathematical results characteristic of relativistic physics.

\section{Hints at new physics}

The study of the previous section seems to suggest that there might be a relationship between friction and
relativistic physics, in the sense that both phenomenologies might follow from a more general theory. The idea would be to develop a theory in which bodies are affected by a universal friction, and in which mass varies according to Reye's law.

The CMB radiation suggests that some privileged frame might exist after
all, a frame that might exert some sort of friction on moving bodies. This friction might be extremely small but still have consequences for the formal structure of the theory.

Let us pause to recall that Newtonian Mechanics is based on the principle of Galileian invariance. As a result, the spacetime of Newtonian mechanics is a 4-dimensional affine space $M$ endowed with a time projection $\pi: M\to T$; the space $S$ being the model for the fibers $S_t=\pi^{-1}(t)$.

Actually, most authors develop Newtonian mechanics as Newton originally did, namely by working with a spacetime that splits as a product  of space and time: $M=S\times T$. This is actually the mathematical structure of an Aristotelian spacetime with an absolute notion of `space' and hence of `rest' (the wordline $(P,t)$ represents a point at `absolute rest'). This approach to Newtonian mechanics certainly simplifies the mathematics of the theory, though the use of such an Aristotelian spacetime appears as artificial and unsatisfactory: the absolute space, which is formally present, is actually unobservable as the laws of Physics, with their Galileian invariance, do not allow for its determination.

It is interesting to observe that the original Aristotelian mechanics, with its  spacetime $S \times T$, was epistemologically more satisfactory than Newtonian mechanics in the above simplified form. Indeed, Aristotle imposed that bodies not exerted upon by forces come ultimately to rest in the absolute frame. The absolute space of Aristotelian mechanics is therefore observable.

It is often claimed that Aristotle, quite naively, did not realize that the natural state of motion of bodies is that of uniform motion, as he did not recognize the role of friction forces.
However,   this is not a fair historical account. As Dugas
\cite{dugas55} points out, Aristotle was very well aware that
probably, in vacuum, a body would have moved in uniform motion
indefinitely. However, from that he inferred that absolute vacuum is
impossible. In other words, it is not because Aristotle did not
recognize the role of friction that he did not arrive at the
 relativity principle, but rather because he regarded the principle that
every body should come to absolute rest as more fundamental than the
relativity principle. Indeed, some kind of deceleration is necessary
in order to give physical observability to the concept of absolute
space which Aristotle was not prepared to abandon. Stated in another way, Aristotle regarded as more  important the spacetime structure $S\times T$, and hence the absolute space idea, rather than the relativity principle. For this reason he had to assume the ultimate deceleration of all bodies so that the  absolute rest frame could become observable.


We see that while in Galileian mechanics friction has a negative connotation, as it spoils the uniform motion of bodies, in Aristotelian mechanics it has a positive connotation, has it is the main element that makes the absolute space observable and hence the very concept of absolute space epistemologically well posed.

A theory embodying a universal friction is naturally a theory that involves some absolute reference frame and conversely. Our sought theory is then  going to share these features and additionally is going to interpolate between Special Relativity and
Aristotelian mechanics.

So our idea is to explore a theory with the following features
\begin{itemize}
\item[(a)]  With respect to the previous treatment of the free block, it lives in one more space dimension, i.e.\ the surface of friction is now a volume.
The normal force is, so to say, in the extra-direction, but we are concerned
only to what happens in the observable 3-space.
\item[(b)] Friction acts over all bodies as in Eq.\ (\ref{jhh})  via a coefficient $a_f$ that is universal and
unrelated to the chemical composition of the matter involved. Save for light, all bodies come ultimately to rest with respect to an absolute reference frame $S_0$, their mass $m$ converging to the their absolute rest mass $\tilde m$.
\item[(c)] Mass is lost as in Eq.\ (\ref{jhg}) and Reye's constant $c$ is identified with the speed of light.
\end{itemize}
Mathematically it develops as a straightforward generalization of the previous theory for the moving block, where the equations there obtained are reinterpreted as  applying to general bodies in the new theory. Of course, by proceeding in this way we shall miss a
field theoretical description. Still this approach might give insights into a theory sharing the above features.

\begin{figure}
\centering
\includegraphics[width=10cm]{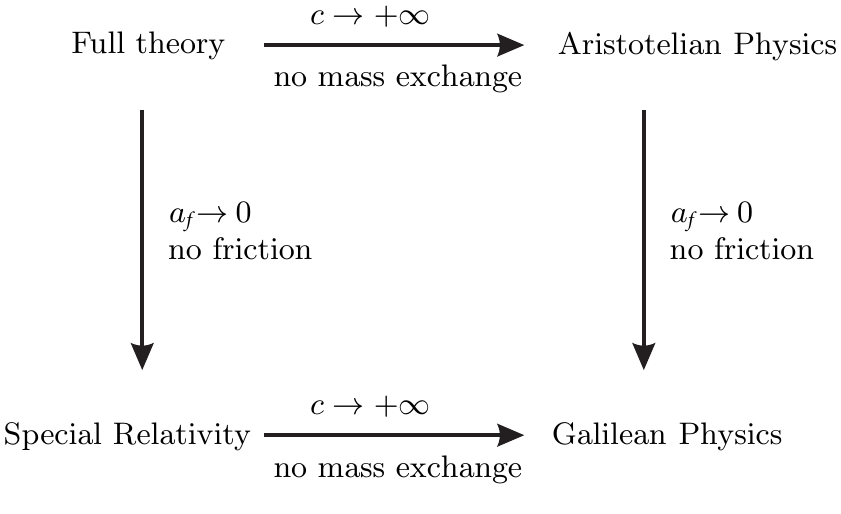}
\caption{Limits of the theory.} \label{com}
\end{figure}

The theory  depends on 2-parameters $a_f$ and $c$ and, as mentioned, shall have two special limits, which are:

\begin{itemize}
\item[SR:] For $a_f\to 0$ friction goes to zero and the absolute space ceases to exist (observe that for $a_f>0$ we can recognize the absolute space as eventually all bodies come to rest with respect to it). In this limit the natural state of bodies is that of uniform motion (relativity principle), however, since the relativistic mass formula (\ref{ngx}) does not depend on $a_f$ and is preserved in the limit, one expects to recover Special Relativity rather than Galileian Relativity.
\item[AM:] For $c\to \infty$ we have by Reye's law $\dot m=0$, masses are constant in time but there is still friction and the theory reduces to Aristotelian mechanics with a universal friction.
\end{itemize}
If from $SR$ we let $c\to \infty$ we naturally obtain Galileian mechanics ($GM$), and similarly, from $AM$ with $a_f\to 0$ we obtain also Galileian mechanics as bodies no more slow down to absolute rest. We arrive at the diagram of Fig.\ \ref{com}.

Let us show that the Lorentz transformation can be generalized so as to depend on two parameters $c$ and $a_f$ and so as to satisfy this scheme. Of course, we are not looking for a group of transformations as the absolute rest frame $S_0$ will be privileged. The group property should be  recovered only for $a_f\to 0$, namely when the privileged frame becomes unobservable. We shall see in a moment that this limit group is indeed the Lorentz group.

 So, let $S_0$ be the absolute frame that parametrizes events with coordinates
$(t_0, x_0, y_0, z_0)$, and let $S$ be a moving frame of coordinates $(t,x,y,z)$ whose origin
moves on the abscissa of $S_0$.

The motion of the origin of $S$ satisfies Eqs.\ (\ref{kde}), (\ref{kdr}), which shows that $S$ will come to rest with respect $S_0$ at time $ t_f=\frac{c \varphi}{a_f}$, its velocity at $t=0$ being $c \tanh \varphi$.

The change of frame
for $t_0< t_f$ is\footnote{For shortness, we omit in the transformation the equations $y=y_0$, $z=z_0$, that should nevertheless be included here as well as  in Eqs.\ (\ref{two}), (\ref{jom}), (\ref{fin}, (\ref{fal}).}
\begin{equation} \label{one}
\begin{split}
t&=-\frac{c}{a_f} \arcsin \tanh (\varphi-\frac{a_f}{c} t_0)+\frac{c}{a_f} \arcsin \tanh (\varphi),\\
x&= \cosh (\varphi-\frac{a_f}{c} t_0)\Big(x_0+\frac{c^2}{a_f} \log \frac{\cosh (\varphi - \frac{a_f}{c} t_0)}{\cosh \varphi}\Big).
\end{split}
\end{equation}
It is such that $(0,0,0,0)\to(0,0,0,0)$, namely the frame origins meet when the respective clocks sign zero time. As mentioned, $\varphi$ is an integration constant related to the velocity of the frame $S$  with respect to $S_0$ at the instant when the origins meet  (this velocity is $c \tanh \varphi$). Setting $x=y=z=0$, namely studying the motion of the origin of $S$, one easily obtains $x_0(t)=-\frac{c^2}{a_f} \log \frac{\cosh (\varphi - \frac{a_f}{c} t_0)}{\cosh \varphi}$, which is of the form (\ref{kdr}), with the integration constant $x_f$ chosen so that $x_0=0$ for $t_0=0$.

For $t_0 \ge t_f$, $S$ is at rest with respect to $S_0$ and the transformation  is just:\\
\begin{equation} \label{two}
\begin{split}
t&=t_0,\\
x&= x_0-\frac{c^2}{a_f} \log \cosh \varphi.
\end{split}
\end{equation}
Having established the compatibility of this transformation with our previous formulas, let us study its limits.

\subsection{Aristotelian limit, $c\to \infty$}
Let us first consider the (Aristotelian) limit $c\to \infty$, which  corresponds to friction without mass loss (cf.\ Reye's law). Since the relative velocity when the origins meet is $v=c \tanh \varphi$, this limit should be taken while taking also the limit $\varphi \to 0$, and in such a way that $c \varphi \to v$ for some constant $v$ to be interpreted as the initial relative velocity.
In this limit from (\ref{one}) we easily get
 for $t_0<v/a_f$
\begin{equation} \label{jom}
\begin{split}
t&=t_0, \\
x&=x_0 - \big(v t_0- \frac{1}{2} a_f t_0^2\big) .
\end{split}
\end{equation}
Observe that this is precisely the Galileian transformation from $S_0$ to  a reference frame that is slowing down with acceleration $a_f$.
For $t_0\ge v/a_f$, the frame has come to rest with respect to $S_0$. Indeed, we get taking the limit of (\ref{two})
\begin{equation} \label{fin}
\begin{split}
t&=t_0, \\
x&=x_0-\frac{1}{2} \frac{v^2}{a_f} .
\end{split}
\end{equation}
In other words we obtain the transformation of  Aristotelian mechanics in which all bodies undergo a  universal deceleration till reaching absolute rest.

\subsection{Special Relativity limit, $a_f\to 0$}

Let us take the limit $a_f\to 0$ in Eq.\ (\ref{one}). Observe that $t_f= \frac{c \varphi}{a_f}\to \infty$, thus we have just  to consider the case $t_0< t_f$.
Let us set $v:=c \tanh \varphi$, $\beta:= v/c$, then we obtain from (\ref{one}) in the limit
\begin{equation} \label{fal}
\begin{split}
t&=\gamma^{-1}(\beta)\, t_0, \\
x&= \gamma(\beta)\, \left(x_0 - v t_0\right) .
\end{split}
\end{equation}
This is the {\em Tangherlini transformation}. It is equivalent to the Lorentz transformation of Special Relativity, the only difference is a resychronization of clocks in the moving frame $S$
\begin{equation}
t_{SR}=t-\frac{v}{c^2} x, \qquad x_{SR}=x.
\end{equation}
The Tangherlini transformation \cite{tangherlini09,malykin09} has been much studied up to this day \cite{guerra06,deabreu08,burde18}. Tangherlini's work, popularied by Mansouri and Sexl \cite{mansouri77}, was anticipated by Scott-Iversen \cite{scott-iversen44} and other authors working on non-standard synchronizations in Special Relativity.

Clocks in Tangherlini's formalism  are set up in such a way that the same notion of simultaneity becomes  shared by all the frames. As a consequence, the speed of light is not isotropic in $S$ (notice that the one-way speed of light depends on the synchronization adopted, as it requires two separate clocks for its measurement, while the two-way speed of light remains unaltered, see \cite{minguzzi02,minguzzi02d}). Of course, in Tangherlini's formalism such a resychronization does not alter the Special Relativity physics, just its coordinate description.

\section{Comparison with observations}

The objective of this section is to  explore the consequences of the  above theory for cosmology. We will show that it predicts the
Pioneer anomaly and the Hubble law, where the latter is explained
through a kind of universal tired light mechanism. Couriously,
the model naturally explains the observed coincidence between the
Hubble constant and the Pioneer anomalous acceleration.

More generally, this model
seems to be able to explain many  odd phenomena that have been
observed for accelerations smaller or of the order of a certain
critical value $a_f$, although  it is not able to explain other observations like  the supernova light curves. Still, these results hint at the possible role of friction in the  resolution of some puzzles of modern cosmology.

\subsection{Pioneer anomaly}
A first obvious consequence of our model is that free bodies with small velocities compared to $c$ should
show a deceleration of magnitude $a_f$ in the absolute reference
frame.

Indeed, any body should decelerate with an acceleration of magnitude
$a_f$. The Pioneer spacecrafts do present an unmodeled acceleration
approximately directed towards the Sun of magnitude
\cite{Anderson:2001sg}
\[
a_P=(8.74 \pm 1.33)\times  10^{-10} m/s^2.
\]
Unfortunately,  according to a recent study, the anomalous
acceleration seems to be non collinear with the spacecraft velocity
\cite{turyshev11}, and can be accounted for by thermal radiation
\cite{turyshev12}.

Also the friction model  predicts a deceleration of the planets of the
solar system, which decreases the radius of their orbits and hence
implies an increase in the modulus of their velocity \cite{parkyn58}
(in the end $\dot{v}=a_p$, that is, the sign that one would naively
expect gets inverted under Coulomb friction). However, this effect
is quite small and can be cancelled by the decrease of the Sun mass
due to solar wind \cite{krasinsky04}.

\subsection{Hubble law}
Let us now consider the distribution of galaxies. Because of friction, if they are close, say at the level of the local group,  they
are almost at rest. Nevertheless, by a
tired light effect the Hubble law still holds. Indeed, a photon sent
at time $t$ from a galaxy and received at time $t+\Delta t$ in the
Milky way undergoes a redshift (see Eq. (\ref{bsc}))
\[
1+z=\frac{E(t)}{E(t+\Delta t)}=e^{\frac{1}{c} \,a_f\, \Delta t}\simeq
1+ \frac{1}{c} \,a_f \,\Delta t.
\]
If this tired light explanation for the  Hubble law is correct then
we must find that the observed value
\[
a_H:= cH=(6.9\pm 0.9) \times 10^{-10} m/s^2
\]
(if $H = (72\pm 8) (km/s)/Mpc$)  coincides with $a_f$, and if the
above explanation for the Pioneer anomaly is correct we must expect
$a_f=a_P$ and in the end
\[
a_H=a_P.
\]
The above figures seem to confirm this prediction. This equality has
been noticed by many authors \cite{Anderson:2001sg}, but this
appears to be the simplest model which accounts for it. This explanation is the more striking as it appears to be compatible with Special Relativity.

We obtained a kind of tired light explanation of the Hubble law, however, contrary to
usual tired light theories, it does not assume that the loss of energy is due to scattering with diffuse interstellar matter. The latter assumption would imply a modification in the direction of the photon, so that any galaxy would be seen as blurred and indefinite, contrary to observations. The universal friction mechanism preserves the direction of the photon and so avoids this problem.

Unfortunately,  tired light explanations of the Hubble law cannot
account for the supernova light curves \cite{leibundgut96}.
Perhaps the Hubble law could be due to two mechanisms, namely tired light for sufficiently close cosmological objects (with the idea that close objects belong to the same vacuum state, the  vacuum state playing the role of the hard body in the mechanical analogy) and cosmic expansion for far away objects (as they belong to different vacuums states, different states diverging from each other). In other words, in the empty region between two vacuum patches there would be no friction and so the dynamics would be completely relativistic (recall that the theory becomes relativistic for $a_f\to 0$, and by Eq.\ (\ref{bsc}) there is no tired light effect in that limit). There would still  be a redshift effect but due to the relative motion of the vacuum states.
In general one would not expect these two redshift effects to match precisely so giving different Hubble constants depending on whether observations are local or cosmological. This tension has indeed been observed and is the subject of current debate.

If the Hubble law
is at least locally due  to tired light, then it should be expected to hold for close
celestial objects as well. The expanding universe theory predicts
the Hubble law at length scales which are well beyond the scale of
homogeneity for which a Friedmann-Robertson-Walker  approximation of the cosmological
metric would make sense. A well known puzzle in cosmology is the
``coldness of the Hubble flow'' namely the observation that the
Hubble law holds unreasonably well at the scale of the local group (1-10Mpc) with a
velocity scatter with respect to the Hubble flow which is very small
(40 km/s), although at that scale the matter distribution is very
clumped \cite{baryshev01}. A local tired light model for the Hubble law
accounts for this observation quite easily, for according to this
explanation the local Hubble law does not depend on an expansion dynamics.

\section{Conclusions}

Inspired by experimental evidence on how friction manifests itself in the wear of bodies, I introduced an absolute space theory in which mass decreases
proportionally to the work done by a universal friction force. I showed that mass depends on velocity as in Special Relativity and that there is a characteristic velocity $c$ which is
insensitive to friction.
Bodies with velocity smaller than $c$
decelerate till they reach a status of absolute rest. Bodies with
velocity $c$ preserve their velocity while they lose mass. I argued
that in the limit of vanishing friction, $a_f\to 0$, the theory
becomes equivalent to Special Relativity (in its Tangherlini coordinate description) as the underlying
absolute space becomes unobservable. This result shows that Special Relativity, often regarded as incompatible with the concept of absolute  frame, could have been suggested  by more detailed studies of the interaction of bodies with such a frame.

I ended the paper with the exploration of some predictions of this theory. Qualitatively, and sometimes also quantitatively, it seems to be able to account for some puzzling cosmological observations. The offered explanations are non-traditional and certainly require further details and clarifications. Still the results of this work suggest that friction might have indeed an important role in cosmology, a role that in my opinion should be explored further.

\section*{Acknowledgments}
I previously posted the work arXiv:1412.0010 that presented some of the arguments of this paper.



\providecommand{\newblock}{}

\end{document}